\documentclass[11pt,prd,showpacs,nofootinbib,oneside,hidelinks,a4paper,english]{revtex4-2}
\usepackage{graphicx}
\usepackage{amssymb}
\usepackage{amsmath}
\usepackage{graphics}
\usepackage{dcolumn}
\usepackage{bm}
\usepackage{color}
\usepackage{epstopdf}
\usepackage{lipsum}
\usepackage{hyperref} 
\usepackage{ulem}

\begin{document}

\newcommand{\blue}[1]{\textcolor{blue}{#1}}
\newcommand{\ans}{\blue}

\title{On Negative Mass, Partition Function and Entropy}

\author{S. D.  Campos\footnote{sergiodc@ufscar.br}}
	
\address{Applied Mathematics Laboratory-CCTS/DFQM,
Federal University of S\~ao Carlos, Sorocaba, CEP 18052780, Brazil}
        
\begin{abstract}
This work examines some aspects related to the existence of negative mass. The requirement for the partition function to converge leads to two distinct approaches. Initially, convergence is achieved by assuming a negative absolute temperature, which results in an imaginary partition function and complex entropy. Subsequently, convergence is maintained by keeping the absolute temperature positive while introducing an imaginary velocity. This modification leads to a positive partition function and real entropy. It seems the utilization of imaginary velocity may yield more plausible physical results compared to the use of negative temperature, at least for the partition function and entropy.
\end{abstract}

\date{\today}
\maketitle

\section{Introduction}
The current laws of physics were formulated under the assumption that the observed mass in the universe is always positive. Nevertheless, certain key equations in physics yield results that suggest the possibility of negative mass. This raises a fundamental question: Why haven't we detected this type of matter? Although experimental evidence is lacking, there are indications that the presence of negative matter could be indirectly inferred through its physical effects. One compelling example is the concept of negative mass being associated with dark energy, which explains the observed acceleration of the universe (see \cite{H.Socas-Navarro.A.A.626(A5).2019} and references therein). 

From a naive perspective, mass is often defined as the amount of matter contained within a given volume. However, in the realm of physics, various conceptual frameworks allow for more nuanced definitions\footnote{A collection of definitions can be found in the Appendix {\it On Relativistic Mass} (written by V. Petkov) in the book of A. Einstein, {\it Relativity}, (Minkowski Institute Press, Montreal, 2018); or, in E. V. Huntington, {\it Bibliographical Note on the Use of the Word Mass in Current Textbooks}, The Amer. Math. Month. \textbf{25}(1), 1 (1918).}, as Max Jammer points out in his book \cite{M.Jammer.book.1}. Classical physics, for instance, is built upon the assumption that these definitions should hold even for negative mass. 

The concept of negative mass in physics is not new, and its existence has been a subject of investigation since at least the end of the nineteenth century \cite{A.Foppl.Sitz.Bayer.Akad.Hiss.Munchen.27.6.1897,A.Schuster.PotentialMatter.AHoliday.Dream.Nature.58(1503).367.1898}. Over the years, numerous studies have explored the possibility of the existence of such an entity \cite{J.M.Luttinger.OnNegativeMassintheTheoryofGravitation.GravityResearchFoundation.UniversityofWiscosin.Madison.Wis.1951,J.Tiomno.Nuovo.Cimento.1.226.1955,J.J.Sakurai.Nuovo.Cimento.7.649.1958,H.Bondi.Rev.Mod.Phys.29.423.1957,P.Morrison.American.J.Phys.26.358.1958,D.Matz.F.A.Kaempffer.Bull.American.Phys.Soc.3.317.1958,L.I.Shift.Proc.Nat.Acad.Sci.45.69.1959,C.de.Beauregard.Comptes.Rendus.252.1737.1961,F.Winterberg.Nuovo.Cimento.19.186.1961,B.Hoffmann.International.Conference.on.Relativistic.Theories.of.Gravitation.(KingsCollege.London).Vol.2.1965,Ya.P.Terletsky.International.Conference.on.Relativistic.Theories.of.Gravitation(KingsCollege.London).Vol.2.1965,A.Inomata.D.Peak.Nuovo.Cimento.63.132.1969,R.deA.Martins.Lett.Nuovo.Cimento.28.8.265.1980,W.B.Bonnor.Gen.Relat.Grav.21(11).1143.1989,R.T.Hammond.arXiv.1308.2683.grqc}.

In general relativity, the seminal approach is due to Bondi, which divides the concept of mass into three entities depending on their measurement origin \cite{H.Bondi.Rev.Mod.Phys.29.423.1957}. The inertial mass, $m_i$, is the one that suffers the effects of inertia. The passive gravitational mass, $m_p$, is the one that feels the force generated by the gravitational potential. The last one, the active gravitational mass, $m_a$, generates the gravitational potential felt by $m_p$. The third law conduces to $m_p=m_a=m$, and the equivalence principle leads to $m_i=m$.

As well known, it is an empirical observation that positive matter attracts positive matter. It is assumed, then, that positive matter attracts {\it all types of mass}, including negative one. Certainly, if such a scenario were to occur, it is plausible an analogy between charge and mass, and it seems reasonable that negative matter should repel negative matter as well as {\it all types of matter}. However, due to the equivalence principle, the positive and negative mass should be accelerated in the same direction when immersed in a gravitational field \cite{Ya.P.Terletskii.book_1}.  

Indeed, the interaction between positive and negative matter can lead to peculiar outcomes in physics \cite{R.T.Hammond.arXiv.1308.2683.grqc}. In the present work, I assume that the laws of classical physics hold for negative mass. Considering this, one studies the partition function and entropy associated with this kind of matter. Some interesting results emerge from the approach considered here. 

This paper is organized as follows. Section \ref{sec:intro} presents some considerations about negative mass. Section \ref{sec:tempvel} introduces some concepts about negative absolute temperature and imaginary velocity. In Section \ref{sec:partentro}, one presents the partition function and entropy for negative mass. The discussion is left for Section \ref{sec:final}.

\section{Positive and Negative Mass Interaction}\label{sec:intro}
In accordance with Bonnor \cite{W.B.Bonnor.Gen.Relat.Grav.21(11).1143.1989}, one assumes here that the equivalence principle, general relativity, and electromagnetism hold for positive and negative mass separately. Considering this, notice that in Newtonian physics the concept of a center-of-mass coordinate system for $n$ particles with positive mass is written as (only for $x$ here)
\begin{eqnarray}
X_{CM}=\frac{m_1x_1+m_2x_2+...+m_nx_n}{m_1+m_2+...+m_n},
\end{eqnarray}

\noindent where $X_{CM}$ stands for the $x$-coordinate in the center-of-mass system, and each $m_j$, $j=1,2,...,n$ represent a mass. On the other hand, considering a system composed of $n$ particles possessing only negative mass, one has the same result as can be easily verified. 

The problem arises when one considers a system comprising $n$ particles with both positive and negative mass. In this case, we do not have a trivial result as above. Indeed, suppose one has $i=1,...k-1$, $1\leq k\leq n$, positive mass particles. Then, one writes the center-of-mass system as
\begin{eqnarray}
X_{CM}=\frac{m_1x_1+m_2x_2+...+m_kx_k+m_{k+1}x_{k+1}+m_{k+2}x_{k+2}+...+m_nx_n}{m_1+m_2+...+m_k+m_{k+1}+m_{k+2}+...+m_n}.
\end{eqnarray}

This result implies the center-of-mass system can only exist in Newtonian physics for positive and negative mass if the sum $m_1+m_2+...+m_j+m_{k+1}+m_{k+2}+...+m_n\neq 0$. Assumes that $m_1+m_2+...+m_k$ is the positive mass contribution while $+m_{k+1}+m_{k+2}+...+m_n$ is due to the negative mass. Therefore, the center-of-mass $X_{CM}$ can only exist if $+m_{k+1}+m_{k+2}+...+m_n=-\gamma (m_1+m_2+...+m_k)$, $\gamma>0$, resulting in
\begin{eqnarray}
X_{CM}=\frac{m_1x_1+m_2x_2+...+m_kx_k+m_{k+1}x_{k+1}+m_{k+2}x_{k+2}+...+m_nx_n}{(1-\gamma)(m_1+m_2+...+m_k)}.
\end{eqnarray}

Hence, a system consisting of $n$ particles with positive and negative mass exhibits a well-defined center-of-mass coordinate only if the sum of positive and negative mass is non-null. For a simple pair composed of two masses $m_1>0$ and $m_2<0$, the above result implies the center-of-mass coordinates exist only if $m_2=-\gamma m_1$, $\gamma>0$.

The explanation above highlights how negative mass can introduce conceptual challenges that extend to well-defined quantities in classical physics. From hereafter, one considers a system containing only either positive or negative mass particles.

\section{Negative Temperature and Imaginary Velocity}\label{sec:tempvel}

\subsection{Negative Absolute Temperature}

As well-known, for a thermal bath of an ideal gas, the probability of occupation of the state $i$ decreases exponentially with energy since \cite{S.Braun.J.P.Ronzheimer.M.Schreiber.S.S.Hodgman.T.Rom.I.Bloch.U.Schneider.Science.339(6115).52.2013}
\begin{eqnarray}
    P_i\propto e^{-\frac{E_i}{k_BT}},
\end{eqnarray}
\noindent where $k_B$ is the Boltzmann constant, $T$ is the absolute temperature\footnote{Hereafter, $T$ referees always to the absolute temperature (positive or negative).}, and $E_i$ is the energy at the state $i$. Consequently, in this scenario, the occupation probability of state $i$ decreases as the energy of that state increases. In other words, the higher the energy in the state, the less likely it is to be occupied. Indeed, this reasoning is meaningful only for positive temperatures ($T > 0$). At absolute zero temperature ($T = 0$), all systems tend to their ground state, and the occupation probability of the lowest energy state becomes unity. However, as the temperature increases ($T > 0$), the probabilities of higher energy states increase, leading to a more distributed occupation of states in accordance with the Boltzmann distribution. 

For negative temperature, there must be an upper bound to $E_i$ \cite{N.F.Ramsey.Phys.Rev.103.20.1956,M.J.Klein.Phys.Rev.104.589.1956}. Surprisingly, there are physical systems exhibiting a negative temperature \cite{E.M.Purcell.R.V.Pound.Phys.Rev.81.279.1951,A.S.Oja.O.V.Lounasmaa.Rev.Mod.Phys.69.1.1997,P.Medley.D.M.Weld.H.Miyake.D.E.Pritchard.W.Ketterle.Phys.Rev.Lett.106.195301.2011,S.Braun.J.P.Ronzheimer.M.Schreiber.S.S.Hodgman.T.Rom.I.Bloch.U.Schneider.Science.339(6115).52.2013}. In general, negative $T$ is defined as that one hotter than the positive one, and it could be even infinite \cite{N.F.Ramsey.Phys.Rev.103.20.1956,M.J.Klein.Phys.Rev.104.589.1956,SDCVAOCVM,SDCVAO}. This concept arises from certain systems that exhibit unique energy-level structures, where higher energy states have lower probabilities of occupation compared to lower energy states, as previously discussed. Moreover, it is important to note that in certain physical systems, negative temperatures can be theoretically achievable. These systems have limited accessible energy levels, which means that adding energy to the system increases its entropy. Additionally, negative temperatures can be interpreted as temperatures tending towards infinity on the Kelvin scale, as they correspond to systems with energy distributions favoring high-energy states over lower-energy states.

\subsection{Imaginary Velocity}

Imaginary velocity is a mathematical consequence of the imaginary time definition, which is a handy tool in special relativity and quantum mechanics. As well-known, imaginary time can be defined through the ordinary time $t$ as
\begin{eqnarray}\label{eq:time}
    \tau = it,
\end{eqnarray}
\noindent being $\tau$ orthogonal to $t$. Consequently, $\tau$ forms a non-ordered set of events\footnote{There is no concept of past, present, and future in $\tau$.}, although the events in $t$ can be ordered. The imaginary time can be used to write the line element in the Minkowski spacetime as
\begin{eqnarray}
\nonumber    ds^2=dx_1^2+dx_2^2+dx_3^2-cdt^2,
\end{eqnarray}
\noindent where $dx_4^2=-cdt^2=c(idt)^2=cd\tau^2$. In addition to the utilization of imaginary time, the imaginary velocity can be also introduced in general relativity as a manifestation of the equivalence principle: gravitational force is a consequence of inertial motion. Therefore, one can define the imaginary velocity $v_i=iv$ where $v$ is the usual real velocity. It means that a negative mass particle has an imaginary linear momentum, but it has a measurable kinetic energy. 

In quantum mechanics, imaginary time is used to explain, for example, particle tunneling through vibrating barriers \cite{V.S.Popov.PhysicsofAtomicNuclei.Vol.68.No.4.686.2005}. 

The set of imaginary velocities can not be ordered, i.e. there is no definition of an upper limiting velocity, for example. Conversely, the limiting velocity $v$ for massive or massless particles in general relativity is the velocity of light $c$. 

Despite its non-straightforward physical interpretation, imaginary velocity is used in certain physical systems. For some elastoplastic materials, the occurrence of imaginary velocities is due to a bad choice of boundary conditions \cite{L.J.Sluys.Wave.propagation.localisation.and.dispersion.in.softening.solids.Dissertation.Delft.University.of.Technology.Delft.1992}. In this case, the appearance of imaginary velocity stems from an incorrect selection of geometry for the given problem. 

A black hole emerging in 10 dimensions $p$-branes is linearly unstable to long-wave perturbations along its world-volume \cite{M.Sun.Y-C.Huang.Nucl.Phys.B897.98.2015}. Within the framework of relativistic hydrodynamics, this instability can be viewed as a dynamic instability of the fluid. The emergence of an imaginary velocity for sound waves means the non-propagation of density perturbations. In this context, the presence of such instability is explained as a consequence of the interplay between the relativistic effects and the behavior of the fluid \cite{M.Sun.Y-C.Huang.Nucl.Phys.B897.98.2015}. In this problem, the imaginary velocity is the manifestation of a non-possibility. 

As a last example, the group velocity at the resonance absorption of a dispersive medium may be negative \cite{L.Brillouin.Wave.propagation.and.group.velocity.Academic.Press.New.York.1960}. However, group velocity can be greater than the phase velocity, turning its physical interpretation a hard matter \cite{R.Landauer.Nature.341.567.1989}. At this point, imaginary velocity can confuse.

Here, one adopts a simple interpretation for the imaginary velocity: if it could be measured, then there exists a physical mechanism preventing this measure from being greater than $c$.  

\section{Partition Function and Entropy}\label{sec:partentro}

The existence of negative mass leads to the necessity of embracing unusual assumptions, such as negative temperature and imaginary velocity, which may consequently yield intriguing results, as will be explored further below.

\subsection{Partition Function}

A partition function describes the statistical properties of a physical system in thermodynamic equilibrium. Mathematically speaking, a partition function is a sum of the number of accessible states of a system, where each state is weighted by a convenient positive number representing the accessibility of such state. Considering this, a partition function is a positive quantity, which is a statistical mechanical requirement. 

However, in some situations, it is possible to obtain a negative partition function. The Potts cluster model is commonly used to study phase transitions and critical phenomena \cite{R.B.Potts.Proc.Camb.Phil.Soc.48.106.1952}. For a line of $n$ even vertices, one writes the partition function for this model as \cite{S.C.Chang.J.L.Jacobsen.J.Salas.R.Shrock.J.Stat.Phys.114.763.2004}
\begin{eqnarray}
    Z(T_n,q,v)=q(q+v)^{n-1},
\end{eqnarray}
\noindent where $q-v<0$ and $T_n$ is the temperature at $n$. The parameter $q$ is limited, $0<q<1$, and at a finite physical temperature, one has $Z(T_n,q,v)<0$. To circumvent this problem, it is assumed $n$ odd. 

Also, the tensor renormalization group method \cite{M.Levin.C.P.Nave.Phys.Rev.Lett.99.120601.2007} can lead to a negative partition function value \cite{C.Wang.S.-M.Qin.H.-J.Zhou.arXiv.condmat.1311.6577.2013}, that can be avoided by assuming a significant cut-off parameter or by increasing the precision of the data \cite{Z.Zhu.H.G.Katzgraber.arXiv.1903.07721.cond.mat.dis.nn.2018}. 

The two above examples show the possibility of negative partition functions, despite the lack of physical interpretation. As shall be seen, the existence of negative mass can lead to this kind of partition function. For our purposes, one begins considering a 1-particle system composed of a positive mass particle, writing its semi-classical partition function as
\begin{eqnarray}\label{eq:partition}
    Z_1^+=4\pi V\left(\frac{m}{h}\right)^3\int_0^{\infty}v^2e^{-\frac{mv^2}{2k_BT}}dv,
\end{eqnarray}
\noindent where $V$ is the reservoir volume, and $h$ is the Planck constant. For a system composed of $N$ identical positive mass particles, the above partition function yields the well-known result
\begin{eqnarray}\label{eq:integ}
    Z_N^+=\frac{1}{N!}\left(\frac{2\pi m k_B T}{h^2}\right)^{3N/2}V.
\end{eqnarray}

It is important to stress that, for a positive mass, the convergence of the integral on the r.h.s of Eq. (\ref{eq:partition}) is obtained assuming that $T$ is always positive, and the reference frame is chosen in such a way that $0<v$.   

From now on, one assumes a gas composed of $N$ identical negative mass particles inside $V$. Notice the particles in such a gas repel each other and repel the internal boundaries of the reservoir. It is plausible that given a finite time $t$, the internal spatial configuration of each particle should tend to an almost stable position\footnote{Quantum mechanics require, at least, small orbits around the stable position.}. The parity of the number of particles influences this configuration, meaning that it is not the same for an even or odd number of particles\footnote{For example, if $V$ is a sphere of radius $r$, the final spatial configurations is different for 1 or 2 negative mass particles.}. In the absence of a reservoir, a plausible scenario is that the negative mass particles move outward from a hypothetical center at a finite time. It is important to note that this effect does not occur for positive mass particles since each particle is mutually attracted to the others.

Taking into account the convergence of integral in Eq. (\ref{eq:partition}), one studies the physical effects of the two possible transformations 
\begin{eqnarray}
\nonumber  (i)~~T\rightarrow -T, ~\mathrm{keeping}~ v\\
\nonumber (ii)~~v\rightarrow iv, ~\mathrm{keeping}~ T
\end{eqnarray}

Adopting the transformation $(i)$, the convergence of the integral in Eq. (\ref{eq:partition}) is guaranteed allowing us to write for the 1-particle system
\begin{eqnarray}
    Z_1^-=-\left(\frac{2\pi mkT}{h^2}\right)^{3/2} V,
\end{eqnarray}
\noindent which corresponds to a negative partition function. In general, this result describes a non-physical system. However, from this result, one can calculate the semi-classical partition function for the $N$-particle system 
\begin{eqnarray}\label{eq:neg_1}
     Z_N^-[T]=(-1)^{N}\frac{1}{N!}\left(\frac{2\pi m k T}{h^2}\right)^{3N/2}V=(-1)^N Z_N^+,
\end{eqnarray}
\noindent where one writes $Z_N^-[T]$ only to emphasizes that $T$ is negative. 

Notice the sign of the result shown in Eq. (\ref{eq:neg_1}) depends on the parity of $N$. If $N$ is odd, then $Z_N^-[T]$ is negative; $N$ even implies that $Z_N^-[T]$ is positive. Then, 
\begin{eqnarray}\label{eq:neg_22}
     Z_N^-[T]&=& Z_N^+, ~~N ~~\mathrm{even},
\end{eqnarray}
\begin{eqnarray}\label{eq:neg_223}
     Z_N^-[T]&=&-Z_N^+, ~~N ~~\mathrm{odd}.
\end{eqnarray}

This anomalous partition function is quite similar to the aforementioned Potts model, where the negative partition problem is avoided by the correct choice of $N$ (even here). Notice, however, that the choice of $N$ (even or odd) may be related to the final internal spatial configuration of the particles, which may produce modifications in the partition function. 

For $N$ is even and negative $T$, the spatial configuration leads to a partition function whose physical meaning describes a system where the low energy states are occupied, as usual. In terms of the partition function given in Eq. (\ref{eq:integ}), there is no difference between a system with an even number of negative mass particles and one with positive mass (whatever its number). Therefore, notwithstanding $T$ is negative, there is no influence of it on the partition function for an even number of negative mass particles.

On the other hand, for a $N$ odd and negative $T$, the spatial configuration leads to a partition function whose physical meaning describes a system favoring the highest energy state occupation in accordance with the negative temperature understanding \cite{N.F.Ramsey.Phys.Rev.103.20.1956,M.J.Klein.Phys.Rev.104.589.1956,SDCVAOCVM,SDCVAO}. Thus, assuming $(i)$, whether $N$ is even or odd leads to different final spatial configurations for the particles in the reservoir, thereby altering the occupation of energy levels within the system. 

Notice that if one assumes $m$, $T$, and $v$ positive, then the partition functions Eqs. (\ref{eq:neg_22}) and (\ref{eq:neg_223}) can be obtained using the simple mathematical transformation given below
\begin{eqnarray}\label{eq:trans1}
    -\frac{mv^2}{2k_BT}\rightarrow -\frac{mv^2}{2k_BT}+\ln (-1)=-\frac{mv^2}{2k_BT}+i\pi.
\end{eqnarray}

The transformation (\ref{eq:trans1}) takes the real energy level into the complex domain, resulting in severe modifications to the partition function, despite $m$, $T$, and $v$ positive. For the 1-particle system, the partition function is
\begin{eqnarray}\label{eq:partitionxx}
    Z_1=4\pi V\left(\frac{m}{h}\right)^3\int_0^{\infty}v^2e^{-\frac{mv^2}{2k_BT}+\ln (-1)}dv=-4\pi V\left(\frac{m}{h}\right)^3\int_0^{\infty}v^2e^{-\frac{mv^2}{2k_BT}}dv,
\end{eqnarray}
\noindent where one can see the emergence of the negative partition function. Thus, the introduction of negative $T$ has the same physical effect as a change in the domain of the energy levels: from the real domain to the complex one. Thereby, one can adopt $m=|m|$ and $T=|T|$ in the partition function and study the effects of a negative mass using the transformations (\ref{eq:trans1}). Therefore, the difference between negative and positive mass in the partition function is due to the introduction of the complex domain to the energy occupation levels.  

On the other hand, adopting the transformation $(ii)$, there is no change in the partition function given by Eq. (\ref{eq:integ}), i.e., its physical interpretation is the usual one. To see that is sufficient to replace $v^2$ by $(iv)^2=-v^2$ in Eq. (\ref{eq:integ}). For the $N$-particle system the resulting partition function is just
\begin{eqnarray}
    Z_N^-[v]= Z_N^+.
\end{eqnarray}

In fact, as mentioned earlier, the utilization of an imaginary velocity may imply the existence of a physical mechanism acting when it is measured. As seen above, the sum of the energy occupation level in the domain of imaginary velocity is a real positive quantity. Thus, an imaginary velocity, despite its nonphysical appeal, has a physical influence on the partition function. Therefore, the transformation $(ii)$ does not produce an anomalous partition function. 


\subsection{Entropy}

In this subsection, only the negative mass case is studied. Then, considering the transformation $(i)$ and for $N$ even, one can write the entropy from the partition function given by Eq. (\ref{eq:neg_22}) as 
\begin{eqnarray}\label{eq:entropy_0}
    S^-[T]=k_B\ln Z_N^+ + k_BT\left(\frac{\partial \ln Z_N^+}{\partial T} \right)_V.
\end{eqnarray}

Despite transformation $(i)$, which leads to an anomalous partition function, the result for $N$ even shown in Eq. (\ref{eq:entropy_0}) is the usual entropy for positive mass particles which is in accordance with the discussion of the preceding subsection. Then, if $N$ is even, then a negative $T$ does not change the usual physical interpretation of the entropy.

Conversely, for $N$ odd, one has the partition function (\ref{eq:neg_223}), resulting in the following entropy
\begin{eqnarray}\label{eq:entropy_1}
       S^-[T]=k_B\ln Z_N^++ k_BT\left(\frac{\partial \ln Z_N^+}{\partial T} \right)_V +k_B\ln (-1)^N.
\end{eqnarray}

One can explicitly write for $N$ odd $N=2n-1$, for $0<n\in \mathbb{Z}$, and from Eq. (\ref{eq:entropy_1}), one has
\begin{eqnarray}\label{eq:entropy_2}
\nonumber S^-[T]&=&k_B\ln Z_N^+ + k_BT\left(\frac{\partial \ln Z_N^+}{\partial T} \right)_V +\ln (-1)^{2n-1}=\\
&=&k_B\ln Z_N^+ + k_BT\left(\frac{\partial \ln Z_N^+}{\partial T} \right)_V + ik_B(2n-1)\pi,
\end{eqnarray}
\noindent which represents a complex entropy, where the real and imaginary parts of $S^-[T]$ are, respectively,
\begin{eqnarray}\label{eq:erim}
    \mathrm{Re}S^-[T]=k_B\ln Z_N^+++ k_BT\left(\frac{\partial \ln Z_N^+}{\partial T} \right)_V ~~\mathrm{and}~~\mathrm{Im}S^-[T]= k_B(2n-1)\pi.
\end{eqnarray}

Then, if $N$ is odd and $T$ is negative, the entropy is complex. The complex entropy is, in general, associated with information measures in the context of Shannon entropy \cite{R.F.NalewajskiJ.Math.Chem.54.1777.2016}. The real part of the complex entropy shown in Eq. (\ref{eq:erim}) corresponds to the entropy associated with positive temperature. Therefore, the real part corresponds to the classical entropy for positive mass particles. On the other hand, the imaginary part in Eq. (\ref{eq:erim}) may represent the entropy of the non-accessible energy states, which does not contribute to the total entropy of the accessible states (the real part). 

Now, considering transformation $(ii)$, one has for $N$ even and odd the same result as shown in Eq. (\ref{eq:entropy_0}). Then,
\begin{eqnarray}
      S^-[v]=k_B\ln Z_N^+ + k_BT\left(\frac{\partial \ln Z_N^+}{\partial T} \right)_V,
\end{eqnarray}
\noindent which is in accordance with the assumption that if a negative mass has imaginary velocity, then it has measurable kinetic energy which contributes to the partition function and entropy. Therefore, if $v$ is imaginary, then there is no modification in the usual physical interpretation of entropy.

\section{Discussion}\label{sec:final}

Assuming $(i)$, then the results for the partition function and entropy may depend on the spatial configuration of the negative mass particles inside $V$. In turn, this configuration depends on whether $N$ is even or odd. For an even number of particles, the entropy for the positive and negative mass particles has the same numerical value. On the other hand, for an odd number of negative particles, the resulting entropy is a complex function where the real part is equal to the entropy for the positive mass case while the imaginary part may represent the entropy associated with the non-accessible energy states. 

However, as commented along the text, for positive mass and temperature, the transformation (\ref{eq:trans1}) yields the same results as assuming a negative mass and temperature. Thus, it seems the use of negative temperature generates the same result as moving the energy levels to the complex domain.  

On the other hand, using $(ii)$, the results for the partition function and entropy are the expected ones. Nonetheless, even though it results in an imaginary linear momentum, it gives rise to a real kinetic energy, which has actual physical implications on the partition function and entropy. Cosmological implications are obvious since negative mass particles can alter the total energy of the system, while maybe undetectable.

Transformation $(i)$ implies both the partition function can be negative and entropy complex while the use of $(ii)$ yields both a real kinetic energy and an imaginary momentum, which may create experimental difficulties in detecting this kind of mass using the usual scattering processes. On the other hand, the introduction of interaction between positive and negative mass particles and the use of $(ii)$ may solve this problem. 


In general, an imaginary velocity is associated with particles traveling faster than light in a vacuum, which would lead to a violation of causality. However, it should be stressed that causality is a concept valid for non-imaginary time and velocity, having no physical meaning for imaginary quantities. Then, the eventual measurement of such particles implies the existence of some physical mechanism preventing the violation of causality for real quantities. This physical mechanism may be similar to the wave-function collapse, where the superposition of states is transformed into a measurable classical state.

Suppose one can use $(ii)$ to the negative mass in a system composed of $N$ positive mass particles and $N$ negative mass. If the entropy in this system is an extensive property, then the resulting total entropy is given by
\begin{eqnarray}
    S_{total}=S_N^+ + S_N^- = 2\left[k_B\ln Z_N^+ + k_BT\left(\frac{\partial \ln Z_N^+}{\partial T} \right)_V\right].
\end{eqnarray}

On the other hand, if the entropy in this system is a non-extensive property, the result can be given in terms of the Tsallis entropy \cite{SDCVAO,SDCVAOCVM}
\begin{eqnarray}
\nonumber S_{total}&=&S_N^+ + S_N^- + (1-q)S_N^+S_N^-=\\
\nonumber &=&k_B\ln Z_N^+\bigl[2 + (1-q)k_B\ln Z_N^+\bigr]+\\
&+&k_BT\left[ 2+(1-q)k_B\left(\ln Z_N^++T\left(\frac{\partial \ln Z_N^+}{\partial T} \right)_V\right)\right]\left(\frac{\partial \ln Z_N^+}{\partial T} \right)_V,
\end{eqnarray}
\noindent where the possible interaction between $S_N^+$ and $S_N^-$ is measured by the entropic-index $q$. If the possible measurements of $S_{total}$ tend to linearity in $\ln Z_N^+$, then $q\rightarrow 1$ while for a non-linear behavior if $q<1$ and non-negligible derivative terms. Thus, $q$ may be used to measure the possible interaction between positive and negative mass particles in the system. 

Another interesting question is total entropy production by negative mass particles in the post-inflationary era \cite{RHL.SDC}. The production of negative mass particles during the decaying of some fundamental field deserves attention since the total entropy production should take into account the negative mass contribution. This question is presently under study. 

Recently, physical properties expected from the negative effective mass were realized by cooling $^{87}$Rb atoms trying to clarify the role of the dispersion relation \cite{M.A.Khamehchi.K.Hossain.M.E.Mossman.Y.Zhang.Th.Busch.M.M.Forbes.P.Engels.PRL118.155301.2017}. The use of potential approach as described in Refs. \cite{M.C.Onyeaju.A.N.Ikot.C.A.Onate.H.P.Obong.O.Ebomwonyi.Commun.Theor.Phys.70.541.2018,M.C.Onyeaju.A.N.Ikot.E.O.Chukwuocha.H.P.Obong.S.Zare.H.Hassanabadi.Few.Body.Syst.57(9).823.2016,A.N.Ikot.H.Hassanabadi.N.Salehl.H.P.Obong.M.C.Onyeaju.Indian.J.Phys.89(11).1221.2015} can be useful to understand some of the physical properties observed in this system. This topic is currently being investigated.


\section*{Acknowledgments}

The author thanks to UFSCar.


\end{document}